\documentclass[aps,prb,floatfix,showpacs,twocolumn,groupedaddress]{revtex4}
\usepackage{graphicx}
\usepackage{dcolumn}

\newcommand{\vs}{{\it vs.}}
\newcommand{\ie}{{\it i.e.}}
\newcommand{\al}{{\it et al.}}
\newcommand{\bavs} {BaVS$_3$}

\newcommand{\bq}{
\begin{equation}
}
\newcommand{\eq}{
\end{equation}
}
\begin{document}

\title{Collective Charge Excitations below the Metal-to-Insulator Transition in
\bavs{}}

\author{T.~Ivek}
\homepage{http://real-science.ifs.hr/}
\email{tivek@ifs.hr}
\author{T.~Vuleti\'{c}}
\author{S.~Tomi\'{c}}
\affiliation{Institut za fiziku, P.O.Box 304, HR-10001 Zagreb, Croatia}
\author{A.~Akrap}
\author{H.~Berger}
\author{L.~Forr\'{o}}
\affiliation{Ecole Politechnique F\'{e}d\'{e}rale, CH-1015 Lausanne,
Switzerland}

\date{\today}

\begin{abstract}
The charge response in the barium vanadium sulfide (\bavs{}) single crystals
is characterized by dc resistivity and low frequency dielectric spectroscopy. A
broad relaxation mode in MHz range with huge dielectric constant $\approx 10^6$
emerges at the metal-to-insulator phase transition $T_{\mathrm{MI}} \approx 67$~K,
weakens with lowering temperature and eventually levels off below the magnetic
transition $T_\chi \approx 30$~K. The mean relaxation time is thermally
activated in a manner similar to the dc resistivity. These features are
interpreted as signatures of the collective charge excitations characteristic
for the orbital ordering that gradually develops below $T_{\mathrm{MI}}$ and
stabilizes at long-range scale below $T_\chi$.
\end{abstract}

\pacs{71.27.+a, 71.30.+h, 72.15.Nj, 77.22.Gm}
% 71.27.+a:
%    Heavy-fermion solids, electron states
%    Strongly correlated electron systems
%
% 71.30.+h:
%    Insulator-metal transitions
%    Metal-insulator transition
%    Peierls instability, metal-insulator transitions
%
% 72.15.Nj:
%    Charge-density waves, one-dimensional conductors
%    Collective excitations, one-dimensional conductors
%    One-dimensional conductivity
%    Peierls instability, electronic conduction
%
% 77.22.Gm:
%    Dielectric loss and relaxation
%    Relaxation processes, in dielectrics

\maketitle

\section{Introduction}

A variety of condensed matter systems with reduced dimensionality and strong
Coulomb interactions between charges, spins and orbitals, display rich phase
diagrams with novel forms of ordering phenomena \cite{Fulde93}. These collective,
broken symmetry phases like charge and spin-density waves, charge and orbital
orderings have been the focus of intense scientific activity in recent years.
Additionally, in transition metal compounds $d$-electrons experience competing
forces: Coulomb repulsion tends to localize electrons, while hybridization with
the extended ligand valence states tends to delocalize them \cite{Maekawa04}.
The subtle balance makes these systems an excellent source for studying diverse
metal-to-insulator (MI) phase transitions usually accompanied with drastic
changes in charge, spin and orbital properties. In particular, the influence of
the orbital degeneracy and orbital ordering on electrical transport and magnetic
properties is rapidly becoming a central issue \cite{Khomskii05}. For example,
in the classic transition metal oxide V$_2$O$_3$ X-ray anomalous scattering
results have given a direct evidence that the orbital occupancy plays a central
role in the physics of this system \cite{Paolasini99}. The spatial ordering of
the occupancy of degenerate electronic orbitals was shown to account for the
anisotropic exchange integrals found in the antiferromagnetic insulator phase.
Another outstanding issue is the question of collective excitations in the
orbital ordered state. Although no theoretical model has been worked out until
now, it is expected that the orbital degrees of freedom can either give rise to
novel collective excitations, {\it e.g.}\ orbital waves, or can strongly
renormalize other excitations \cite{Pen97}.

The perovskite-type sulfide \bavs{} represents an exceptional system to study
the aforementioned phenomena and as such has attracted much attention in the
last years. Face sharing VS$_6$ octahedra stack along the crystallographic
$c$-axis and form the VS$_3$ spin chains with one 3$d$ electron per V$^{4+}$
site. The chains are separated by Ba atoms in the $ab$ planes, yielding a
quasi-one-dimensional (quasi-1D) structure. The unit cell at room temperature
(RT) is a primitive hexagonal with two formula units. At 240~K the structure
transforms into orthorhombic, but each chain keeps two equivalent V$^{4+}$
atoms per unit cell. The corresponding two electrons are shared essentially
between two hybridized bands produced by the crystal field splitting: a broad
$A_{1g}$ band derived from $d_{z^2}$ orbitals overlapping along the $c$-axis,
and a quasi-degenerate narrow $E_{g1}$ band originating from $e(t_{2g})$
orbitals with isotropic interactions via V-S-S-V bonds
\cite{Whangbo02,Lechermann05,Lechermann06,Lechermann07}. The filling of these
bands is controlled by the on-site Coulomb repulsion $U$ and local Hund's rule
coupling $J$. The main effect is to bring the occupancies of $A_{1g}$ and
$E_{g1}$ orbital closer to one another, which in the limit of strong
correlations yields the occupancy for each of these orbitals close to
half-filling. The spin degrees of freedom of the localized electrons and the
coupling of conduction and localized electrons make the system extremely
complex. Their interplay results in the MI transition at about 70~K and a
magnetic transition at about 30~K.

In spite of a great deal of experimental efforts, no definite understanding has
been reached yet on the detailed nature of the MI phase transition and the
ground state in \bavs{}. Diffuse X-ray scattering experiments have shown that at
$T_{\mathrm{MI}}$ a commensurate superstructure with the critical wave vector
$\mathbf{q}_c = 0.5\mathbf{c}^\ast$ close to $2k_F(A_{1g})$ sets in, preceded by
a large fluctuation regime extending up to 170~K \cite{Fagot03}. This behavior
is reminiscent of a Peierls transition into a charge-density wave (CDW) state.
However, the nature of the ground state is certainly more complicated since
$2\mathbf{q}_c$ harmonic is found suggesting that the localized $e(t_{2g})$
electrons are also involved in the transition and order below $T_{\mathrm{MI}}$.
Indeed, the related susceptibility, which follows the Curie-Weiss law from RT,
exhibits an antiferromagnetic (AF)-like cusp at $T_{\mathrm{MI}}$ and decreases
below \cite{Mihaly00}. There is no sign of magnetic long-range order down to
$T_\chi \approx 30$~K, where the incommensurate magnetic ordering is established
\cite{Nakamura00}. The next unusual result at $T_{\mathrm{MI}}$ is a structural
transformation from the orthorhombic to monoclinic with internal distortions of
VS$_6$ octahedron and tetramerization of V$^{4+}$ chains \cite{Fagot05}.
The modulation of V distances in the superstructure can be described as a
superposition of a $2k_F$ and a $4k_F$ bond order waves (BOW), the latter
component being primarily responsible for the intensity change of the basic
Bragg reflections. It is noteworthy that the tetramerization might be understood
as an inherent feature of the $2k_F$  Peierls transition, which reveals that a
dimerization gap at high temperatures is negligible and that the broad $A_{1g}$
band could be regarded as effectively one-quarter filled band. Finally and the
most surprisingly, X-ray anomalous scattering at the vanadium $K$-edge revealed
no charge disproportionation in the ground state \cite{Fagot06}. A very intriguing
interpretation has been suggested involving the stabilization of two
out-of-phase CDWs from $d_{z^2}$ and $e(t_{2g})$ electrons, which implies an
orbital ordering {\it via} an out-of-phase modulation of the occupancy of V
sites by the $d_{z^2}$ and $e(t_{2g})$ orbitals.

In this paper, we address these important questions concerning the insulating
state in \bavs{}. Our results present a supporting evidence for an orbital
ordered ground state and that the MI Peierls-like transition is also an orbital
ordering transition. We find that a huge dielectric constant which arises in
close vicinity of the MI transition dramatically decreases on cooling down to
about 30~K and levels off below. We argue that such a behavior might be rather
well explained in terms of an orbital ordering which sets in at the MI
transition and develops the long-range order below the magnetic transition.

\section{Experimental and Results}

dc resistivity was measured between RT and 10~K. In the frequency range
0.01~Hz--10~MHz the spectra of complex dielectric function were obtained from
the complex conductance measured by two set-ups. At high frequencies
(40~Hz--10~MHz) an Agilent 4294A precision impedance analyser was used. At low
frequencies (0.01~Hz--3~kHz) a set-up for measuring high-impedance samples was
used \cite{Pinteric01}. The employed ac signal level of
50~mV was well within the linear response regime. All measurements were done on
single crystals along the crystallographic $c$-axis. The typical crystal
dimensions were $3 \times 0.25 \times 0.25$~mm$^3$. 

An influence of extrinsic effects, especially those due to contact resistance
and surface layer capacitance, was ruled out with scrutiny. In order to
determine the quality of contacts we have performed dc resistance measurements
in the standard 4- and 2-contact configurations. Taking into account the
difference in geometry between contacts in these two configurations, the
4-contact resistance can be scaled and subtracted from the 2-contact resistance
in order to estimate contact resistance ($R_\mathrm{c}$) \vs{}\ sample bulk
resistance ($R_\mathrm{s}$). Our analysis clearly shows that good quality
contact samples whose dielectric response reflects the intrinsic features of
the sample bulk can easily be distinguished from bad samples by featuring the
$R_\mathrm{c}/R_\mathrm{s}$ ratio not larger than 5 in the metallic phase and in
the vicinity of $T_{\mathrm{MI}}$, and $R_\mathrm{c}=0.1R_\mathrm{s}$ in the
whole range of the insulating phase except close to $T_{\mathrm{MI}}$. In
contrast, bad samples displayed $R_\mathrm{c}/R_\mathrm{s}$ ratio of the order
of 100--1000 in the metallic phase and in the vicinity of $T_{\mathrm{MI}}$,
indicating large contact resistances, which concomitantly imply large contact
capacitances throughout the whole measured temperature range. Consequently, the
dielectric response registered in samples with bad quality contacts was a
combination of the sample bulk and contact capacitance influence, the latter
becoming dominant with increasing temperature. In this way, out of nine samples
studied, seven were discarded either due to bad contacts, or due to
exceptionally low RRR at high pressure of 20~kbar. Two remaining good quality
contact samples with high RRR at high pressure have shown qualitatively
the same dielectric response, confirming that the observed response comes from
the sample bulk. In this paper we present and discuss results obtained on one of
these two single crystals.

%%%% Figure 1 %%%%
\begin{figure}
\centering\includegraphics[clip,width=0.8 \linewidth]{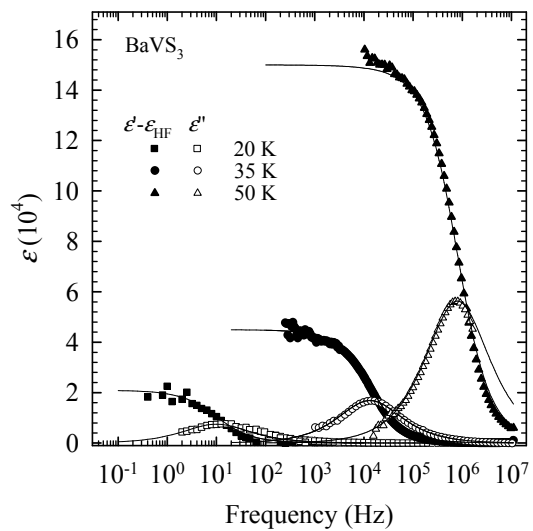}
\caption{Real and imaginary parts of the dielectric function of \bavs{}
measured at three representative temperatures as a function of frequency with
the ac electric field applied along the $c$-axis. The full lines are from fits
by the generalized Debye expression (see Text).}
\label{fig1}
\end{figure}
%%%% Figure 1 %%%%

Fig.\ \ref{fig1} shows frequency dependence of the complex dielectric response
at three selected temperatures. A pronounced dielectric relaxation is observed.
The screened loss peak $\varepsilon''$ centered at $\tau_0^{-1}$ moves toward
lower frequencies and smaller amplitudes with decreasing temperature. The main
features of this relaxation are well described by the generalized Debye
expression $\varepsilon(\omega)-\varepsilon_{\mathrm{HF}} = \Delta\varepsilon / \left[ 1+(i \omega \tau_0)^{1-\alpha} \right]$,
where $\Delta\varepsilon = \varepsilon_0 - \varepsilon_{\mathrm{HF}}$ ($\varepsilon_0$
and $\varepsilon_{\mathrm{HF}}$ are static and high-frequency dielectric
constant, with the latter being negligible), $\tau_0$ is the mean relaxation
time and $1-\alpha$ is the symmetric broadening of the relaxation time
distribution function.

\begin{figure}
\centering\includegraphics[clip,width=1.00 \linewidth]{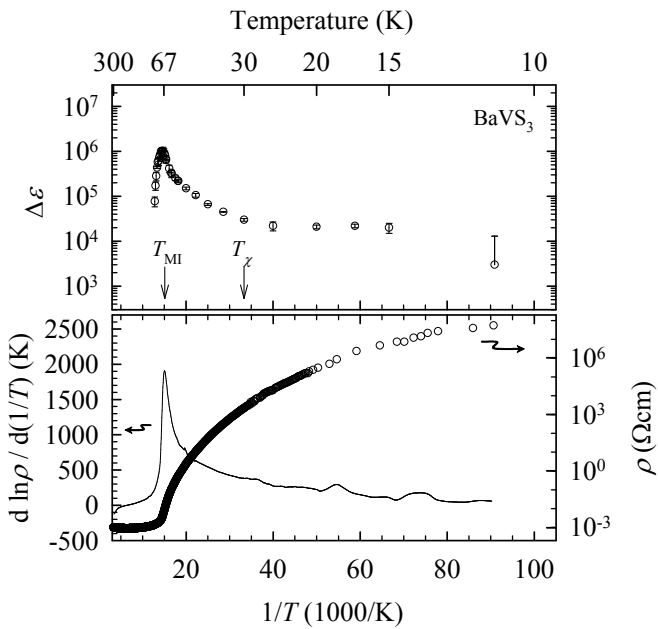}
\caption{Temperature dependence of the dielectric constant 
of the collective mode (upper panel), and dc resistivity and its
logarithmic derivative (lower panel) in \bavs{}. The arrows indicate the
MI and magnetic transition temperature.}
\label{fig2}
\end{figure}

Our results clearly demonstrate that a huge dielectric constant
$\Delta\varepsilon$ is associated with the metal-to-insulator phase transition
(Fig.\ \ref{fig2}). On decreasing temperature, a sharp growth of
$\Delta\varepsilon$ starts in the close vicinity of $T_{\mathrm{MI}}$ and
reaches the huge value of the order of $10^6$ at $T_{\mathrm{MI}} = 67$~K
(Fig.\ \ref{fig2}, upper panel). This $T_{\mathrm{MI}}$ value corresponds
perfectly well to the phase transition temperature as determined in the dc
resistivity measurements, indicated by pronounced peaks at $T_{\mathrm{MI}}$ in
the logarithmic derivative of resistivity (Fig.\ \ref{fig2}, lower panel).

%%%% Figure 2 %%%%
%%%% Figure 2 %%%%

\begin{figure}
\centering\includegraphics[clip,width=1.00 \linewidth]{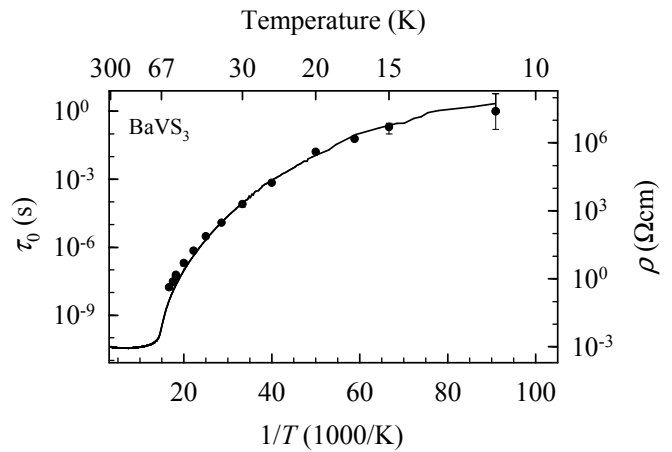}
\caption{The mean relaxation $\tau_0$ of the collective charge mode (points)
and the dc resistivity (full line) in \bavs{} as a function of the
inverse temperature.}
\label{fig3}
\end{figure}

The observed dielectric relaxation would suggest a charge-density wave formation
at $T_{\mathrm{MI}}$ \cite{Gruener88}. The standard model of a deformable CDW
pinned in a non-uniform impurity potential accounts for the existence of two
modes, transverse and longitudinal \cite{Littlewood87}. The former couples to
the electromagnetic radiation and yields an unscreened pinned mode in the
microwave region. Unfortunately, no microwave measurements on \bavs{} have been
done yet. The longitudinal mode couples to an electrostatic potential and mixes
in the transverse response due to the non-uniform pinning resulting in an
overdamped low-frequency relaxation at $\tau_0^{-1}$ due to screening effects.
The relaxation detected in our experiments bears two features as expected in the
standard model. The first is that the relaxation time distribution is
symmetrically broadened, $1-\alpha \approx 0.8$. The second is that the mean
relaxation time $\tau_0$ closely follows a thermally activated behavior similar
to the dc resistivity $\tau_0(T) $ = $\tau_{00} \exp(2\Delta / 2k_\mathrm{B}T) \propto \rho(T)$
(see Fig.\ \ref{fig3}). $\tau_{00} \approx 1$~ns describes the microscopic
relaxation time of the collective mode and the gap $2\Delta \approx 500$~K, as
found in the spectra of the optical conductivity \cite{Kezsmarki06}. The
dissipation can be naturally attributed to the screening due to single particles
originating from the wide $A_{1g}$ band. However, our discovery finds no
consistent explanation within the standard model \cite{Gruener88}, since the
dielectric constant $\Delta\varepsilon$ displays a strong decrease below
$T_{\mathrm{MI}}$ and levels off at temperatures below about 30~K, a behavior
which significantly deviates from the one expected for a CDW condensate density
$n$: $\Delta\varepsilon(T) \propto n(T)$. The decrease of
$\Delta\varepsilon$ on moving between the MI transition $T_{\mathrm{MI}}$ and
the magnetic transition $T_\chi$ is substantial and amounts to two orders of
magnitude. One possible explanation for this discrepancy is the very nature of
the standard model for the response of the conventional CDW to applied electric
fields in which the long-wavelength collective CDW excitation {\it s.c.}\ phason
keeps the prominent role. We remind that this model is worked out for the
incommensurate CDW in a random impurity potential, whereas the CDW in \bavs{}
associated with the observed lattice modulation is commensurate with the order
of commensurability, \ie{}\ the ratio of the CDW and lattice periodicity,
$N = 4$. However, the order of commensurability is not too high to impose the
commensurability pinning and forbid the phason excitations \cite{LRA74}. Indeed,
the experimental observation of the broad relaxation, \ie{}\ $1-\alpha \approx 0.8$
indicates a randomness of the background structure.

%%%% Figure 3 %%%%
%%%% Figure 3 %%%%

Another puzzling feature is brought by the dc electric-field dependent measurements.
In the standard 1D CDW compounds, the collective CDW mode couples to an applied
dc electric field and gives a novel contribution to the electrical conductivity  
\cite{Gruener88}. Our dc electric-field dependent measurements up to fields as
high as 100~V/cm in the temperature range between 15~K and 40~K have only
revealed a negligibly small non-linear conductivity, which emerges from the
background noise.

\section{Discussion}

Our results indicate that the long-wavelength CDW collective
excitations are frozen or strongly renormalized and that the collective
excitations of different kind should be responsible for the observed dielectric
relaxation. In the following we first address other possible causes and then
offer the most plausible scenario for our results. 

First, we verify if the observed behavior of the dielectric constant might be
due to the hopping conduction which is known to arise in disordered systems with low
dimensionality, where Anderson localization yields the conductivity characterized
by variable-range hopping (VRH) Mott law in dc limit, while in ac limit it
follows a power law dependence on frequency. Although \bavs{} may be regarded as
a quasi-1D system, the hopping scenario does not seem realistic for the following
reasons. First is that the frequency marking the onset of the frequency-dependent
transport is known to be roughly proportional to the dc conductivity, \ie{}\ to
the inverse of the dc resistivity \cite{Dyre00}. The dc resistivity of \bavs{}
at RT is about $10^{-3}$~$\Omega$cm and at lowest temperatures it is about $10^5$
to $10^6$~$\Omega$cm. In the diverse systems with dc resistivities of similar
orders of magnitude, the ac conductivity power law is observed only at
frequencies above 1~MHz, whereas below 1~MHz an influence of hopping on
dielectric dispersion is detected only for dc resistivities much higher than
$10^{10}$~$\Omega$cm.\cite{Lunkenheimer03,Vuletic03} Indeed,
in the case of \bavs{} a crude estimate for the crossover frequency yields
values far above the frequency range where the dielectric response was observed
\cite{Vuletic03}. The second result which excludes hopping comes from the observed
optical spectra \cite{Kezsmarki06}. Namely, a simple indication for existence
of a hopping mechanism would be that the optical conductivity is significantly
enhanced in comparison with the dc conductivity, whereas in the case of \bavs{}
the optical conductivity, even at temperatures lower than $T_{\mathrm{MI}}$, is
at best comparable to the dc conductivity. 

Next possible origin of the observed dielectric response might be associated
with a ferroelectric nature of the MI transition. Simple space group
considerations indicate that below the MI transition the structure of \bavs{} is
noncentrosymmetric with a polar axis in the reflection plane containing the VS$_3$
chains. The symmetry of this low-temperature superstructure is {\it Im},
implying that the distortions of the two chains of the unit cell are out of phase
\cite{Fagot05}. Bond-valence sum (BVS) calculations of these X-ray data have
indicated some charge disproportionation at low temperatures. However, it was
recently argued by P.~Foury-Leylekian \cite{Foury07} that the BVS method
overestimated the charge disproportionation due to several reasons: a nonsymmetric
V$^{4+}$ environment, thermal contraction corrections which were not included and
rather imprecise atomic coordinates which were used. Taking all this into account
together with X-ray anomalous scattering result that showed negligible redistribution,
not larger than 0.01 electron below $T_{\mathrm{MI}}$ \cite{Fagot06}, we conclude
that ferroelectricity cannot provide an explanation of the dielectric response
in \bavs{}. 

Finally, we address orbital ordering as a plausible ground state whose
collective excitations might yield the observed dielectric relaxation. First we
list arguments developed by Fagot \al{}\ \cite{Fagot06} who invoked an orbital
ordering associated with the MI phase transition in order to consistently
explain structural data of \bavs{}. Supporting results for an orbital ordering
scenario at $T_{\mathrm{MI}}$ are an almost non-existant charge modulation in the
insulating phase, together with a qualitative structural analysis of the VS$_6$
octahedron
distortions, which reveals an out-of-phase modulation of the occupancy of V
sites by the $d_{z^2}$ and $e(t_{2g})$ orbitals. In particular, dominant
$E_{g1}$ and $A_{1g}$ occupancies are proposed for V1 and V3 sites respectively,
while no definitive preferential occupancy was found for V2 and V4 sites.
Furthermore, recent LDA+DMFT (local density approximation with dynamical
mean-field theory) calculations performed in the monoclinic insulating phase of
\bavs{} have qualitatively confirmed an orbital-ordering scenario showing a
V-site-dependent orbital occupancy and only minor, if any, charge
disproportionation \cite{Lechermann07}. However, these calculations
suggest quite different orbital occupancies: it appears that the (V3,V4) pair
forms a correlated dimer with mixed $A_{1g}$ and $E_{g1}$ occupancy, while the
V1 and V2 ions bear major $E_{g1}$ occupancy and negligible coupling. Finally,
the overall study indicates that although the local environment of the V site
does not change substantially, the electronic structure turns out to be rather
sensitive to change of temperature. The question arises what is the temperature
dependence of the order parameter of orbital ordering which starts to
develop below $T_{\mathrm{MI}}$ and in what way it relates to the magnetic
ordering. $^{51}$V NMR and NQR measurements also suggested an orbital ordering below
$T_{\mathrm{MI}}$ that is fully developed only at $T < T_\chi$ \cite{Nakamura97}.
The magnetic phase transition at $T_\chi$ is preceded by long-range dynamic AF
correlations all the way up to $T_{\mathrm{MI}}$ and this phase bears features
of a gapped spin-liquid-like phase. Mih\'{a}ly \al{}\ \cite{Mihaly00} suggested
that the lack of magnetic long-range order between $T_{\mathrm{MI}}$ and
$T_\chi$ might be the consequence of the frustrated structure of a triangular
array of V chains, which also prevents the orbital long-range order, so that the
long-range spin and orbital orders can develop only well below $T_{\mathrm{MI}}$.
Finally, the AF static order below $T_\chi$ is not a conventional N\'{e}el phase:
an AF domain structure is suggested by the magnetic anisotropy measurements
\cite{Miljak07}. The existence of domains seems to be supported also by the muon
spin rotation ($\mu$SR) measurements, which showed an essentially random
distribution of sizeable static electric fields below $T_\chi$ indicating an
incommensurate or disordered magnetism \cite{Higemoto02}.

Based on the considerations above we suggest the following as the most plausible
scenario. The primary order parameter for the MI phase transition is 1D CDW
instability and this CDW instability drives the orbital ordering {\it via}
structural changes involving a transformation from the orthorhombic to monoclinic
with internal distortions of VS$_6$ octahedron and tetramerization of V$^{4+}$
chains. The orbital order is coupled with the spin degrees of freedom and drives
the spin ordering into an AF-like ground state below 30~K. In other words, the
orbital ordering transition happens at $T_{\mathrm{MI}}$, domains of orbital
order gradually develop in size with lowering temperature (concomitantly their
number diminishes) and the long-range order eventually stabilizes below $T_\chi$,
albeit domain structure persists. In this scenario we
propose that the collective excitations responsible for the observed
features of the dielectric relaxation are short-wavelength ones, like
charge domain walls \cite{note} in the random AF domain structure.
Similar short-wavelength excitations associated with domain structure have
previously been invoked as the origin of dielectric relaxation in diverse
systems \cite{Okamoto91, Pinteric99}.  The relaxation happens
between different metastable states, which correspond to local changes of the
spin configuration. The spin configuration is intimately connected with the
charge and the orbital degrees of freedom as explained above. Since the
dielectric constant is associated with the density of collective excitations,
its anomalous temperature behavior below $T_{\mathrm{MI}}$ indicates that the
relaxation-active number of domain walls decreases with lowering temperature and
eventually becomes well defined below $T_\chi$. In other words, the dynamics of
domain walls becomes progressively more restricted as the temperature lowers and
it becomes constant below $T_\chi$. 

\section{Conclusion}

In conclusion, we demonstrated the appearance of a huge dielectric constant
associated with the metal-to-insulator phase transition in \bavs{} followed by a
dramatic decrease on cooling down to the magnetic transition and leveling off
below. We argue that the collective excitations whose dispersion we detect as
broad screened relaxation modes are not CDW phason excitations; rather they
represent short-wavelength excitations of an orbital ordering, which sets in at
the metal-to-insulator transition and develops the long-range order below the
magnetic transition. Finally, \bavs{} represents a beautiful example of the
intricate interplay between an orbital degeneracy on the one side, and spin and
charge sector on the other side. This interplay needs to be taken into account in order
to understand the origin of the metal-to-insulator phase transition and
low-temperature phases in the transition metal compounds in general. Further work on the
theoretical and experimental fronts is needed to demonstrate directly the
existence of an orbital order in \bavs{} and the associated superstructure.

We thank N.~Bari\v{s}i\'{c}, S.~Bari\v{s}i\'{c}, P.~Foury-Leylekian, V.~Ilakovac,
M.~Miljak and J.~P.~Pouget for useful discussions. This work was supported by the
Croatian Ministry of Science, Education and Sports under Grant No.035-0000000-2836.
The work in Lausanne was sponsored by the Swiss National Science Foundation through
the NCCR pool MaNEP.


\begin{thebibliography}{99}
\bibitem{note}
We note that the domain wall and soliton both stand for short wavelength excitations; however whereas soliton is usually a one-dimensional object, the domain wall is not dimensionally restricted.
\end{thebibliography}
\end{document}